# Intraoperative time out to promote the implementation of the critical view of safety in laparoscopic cholecystectomy: a video-based assessment of 343 procedures

Pietro Mascagni[1, 2], MD; María Rita Rodríguez-Luna[3], MD; Takeshi Urade[4], MD, PhD; Emanuele Felli[5], MD; Patrick Pessaux[5] MD, PhD; Didier Mutter[3, 4, 5], MD, PhD, FACS; Jacques Marescaux[3] MD, FACS, (Hon) FRCS, (Hon) FJSES; Guido Costamagna[2, 6], MD, FACG, FJGES; Bernard Dallemagne[3, 4, 5], MD; Nicolas Padoy[1, 4], PhD

1. ICube, University of Strasbourg, CNRS, IHU Strasbourg, France
2. Fondazione Policlinico Universitario A. Gemelli IRCCS, Rome, Italy
3. Institute for Research against Digestive Cancer (IRCAD), Strasbourg, France
4. IHU-Strasbourg, Institute of Image-Guided Surgery, Strasbourg, France
5. Department of Digestive and Endocrine Surgery, University of Strasbourg, Strasbourg, France
6. Center for Endoscopic Research, Therapeutics and Training (CERTT), Università Cattolica S. Cuore, Rome, Italy

**Corresponding author**:

Pietro Mascagni, MD,

ICube, c/o IHU-Strasbourg,

1, place de l'hôpital,

67000 Strasbourg, France

Phone: +33 3 90 41 36 00

Fax: +33 3 90 41 36 99

Email: p.mascagni@unistra.fr

**Funding**: This study is partially supported by an EAES Research Grant and by French State Funds managed by the "Agence Nationale de la Recherche (ANR)" National Research Agency through the "Investissements d'Avenir" (Investments for the Future) Program under the ANR-10-IAHU-02 (IHU-Strasbourg) grant.

**ABSTRACT**

**Background:** The critical view of safety (CVS) is poorly adopted in surgical practices although it is ubiquitously recommended to prevent major bile duct injuries during laparoscopic cholecystectomy (LC). This study aims to determine whether performing a short intraoperative time out can improve CVS implementation.

**Methods:** Surgeons performing LCs at an academic centre were invited to perform a 5-second long time out to verify CVS before dividing the cystic duct (5-second rule). The primary endpoint was to compare the rate of CVS achievement between LCs performed in the year before and the year after the 5-second rule. The CVS achievement rate was computed using the mediated video-based assessment of two independent reviewers. Clinical outcomes, LC workflows, and postoperative reports were also compared.

**Results:** Three hundred and forty-three (171 before and 172 after the 5-second rule) of the 381 LCs performed over the 2-year study were analysed. After the implementation of the 5-second rule, the rate of CVS achievement increased significantly (15.9 *vs* 44.1 %; $P<0.001$) as well as the rate of bailout procedures (8.2 *vs* 15.7 %; $P=0.045$), the median time to clip the cystic duct or artery (17:26 [interquartile range: 16:46] *vs* 23:12 [17:16] minutes; $P=0.007$), and the rate of postoperative CVS reporting (1.3 *vs* 28.8 %; $P<0.001$). Morbidity was comparable (1.75 *vs* 2.33 % before and after the 5-second rule respectively; $P=0.685$).

**Conclusion:** Performing a short intraoperative time out improves CVS implementation during LC. Systematic intraoperative cognitive aids should be studied to sustain the uptake of guidelines.

**Introduction**

Surgical societies are united[1] in promoting research, education, and quality improvement (QI) initiatives to prevent bile duct injuries (BDIs) since this dreaded adverse event of laparoscopic cholecystectomy (LC) puts a heavy burden on patients' survival[2] and quality of life[3], surgeons' career[4, 5], and health systems' expenditures[6].

Achieving a critical view of safety (CVS)[7] in LC is widely recommended to prevent the visual perceptual illusion which causes 97 % of major BDIs[8], the most severe iatrogenic lesion of the common bile duct. However, the incidence of BDIs in LC remains stable at 0.3 to 1.5 %[2, 9], a rate at least three times higher than the figures commonly reported for open cholecystectomy 30 years ago[10]. In addition, the complexity of BDIs seems to have increased over time, with a trend towards more proximal injuries[11].

The frequently described poor uptake and subjective assessment of formal CVS in surgical practices[12-14] might account for the non-decreasing rate of BDIs[9]. To promote the implementation of CVS, multi-society guidelines on BDI prevention suggested to perform a momentary pause to recall and verify the CVS before clipping and dividing the cystic duct or artery[1] based solely on the opinion of experts.

It has been hypothesised that performing an intraoperative time out could serve as a procedural cognitive aid to recall and apply essential safety measures such as CVS, in the same way as the implementation of the surgical safety checklist[15] serves as a cognitive aid in the perioperative setting[16].

The aim of the present before and after QI study was to evaluate the effect of performing a 5-second long intraoperative time out before the division of the cystic duct through a video-based assessment (VBA)[17] of a large series of LCs.

**Methods**

This QI study was approved by the local medical research and ethical committee (*CE-2020-178*). The study follows a before and after design and it is reported according to the Standards for QUality Improvement Reporting Excellence (SQUIRE) guidelines[18].

Participants

To facilitate QI research and education, patients undergoing surgery in the Department of General, Digestive, and Endocrine Surgery of the Nouvel Hôpital Civil (Strasbourg, France) are routinely asked to give their consent for data recording. Prospectively collected data of patients over 18 years of age and undergoing a LC for benign conditions between November 2017 and November 2019 were analysed retrospectively.

Quality improvement intervention

As part of an institution-wide initiative to promote the implementation of best practices to ensure a safe LC[19-23], two authors of the study (PM and NP) were invited for a short presentation during the morning surgical staff meeting. During this brief intervention, the authors asked surgeons to verify CVS achievement in a 5-second long intraoperative time out before clipping or dividing the cystic duct or artery. To foster attention and reinforce the concept[24], the first operators were asked to indicate and verbalise CVS criteria to their assistant during the 5-second time out. This procedure was called the "5-second rule" (Figure 1). Surgeons were not exposed to further education on CVS or interventions on LC during the study period.

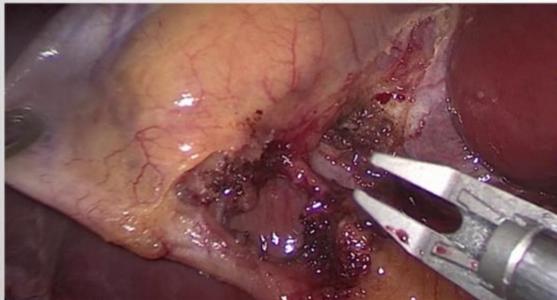

**Figure 1.** The 5-second rule QI intervention. A slide used to explain the concept of the 5-second rule intraoperative time out. A parallelism was drawn with aviation given the extensive use of cognitive aids such as checklists and time outs to improve the safety of flights.

<u>Video-based assessment</u>

Laparoscopic videos of LC procedures performed the year before and the year after the 5-second rule were independently reviewed by a surgical trainee (PM) and a hepatobiliary pancreatic surgeon (TU) with over 300 LCs of experience. Reviewers timestamped the beginning, the first application of a clip in the hepatocystic triangle, and the end of each procedure. Following extensive training, the two independent reviewers assessed the achievement of CVS criteria according to a previously validated binary method[21]. The CVS was defined as the view of only 2 tubular structures, the cystic duct and artery, entering the gallbladder (2-structure criterion, C1), a hepatocystic triangle well dissected from adipose and connective tissue (hepatocystic triangle criterion, C2), and the separation of the lowest part of the gallbladder from the cystic plate (cystic plate criterion, C3)[7]. In the event of disagreement on CVS assessment, mediation was conducted by a third study author (BD) with over 3500 LCs of experience. Finally, video reviewers annotated whether an intraoperative cholangiogram (IOC) was performed and if surgical operators bailed out to a fundus-first cholecystectomy, subtotal cholecystectomy or converted/aborted the laparoscopic procedure.

Outcomes and statistical analysis

The primary endpoint of the study was to compare the rate of CVS achievement the year before and the year after the implementation of the 5-second rule. The CVS achievement rate was computed using the mediated VBA of CVS criteria, after exclusion of bailout (fundus-first or subtotal cholecystectomy), converted or aborted procedures.

Secondary endpoints aimed to compare the LC procedures performed before and after the 5-second rule regarding clinical outcomes, bailout, IOC, operating times, and postoperative CVS reporting.

Patients' baseline characteristics, operative reports, and clinical outcomes data were collected from electronic medical records whereas all other endpoints were annotated on LC videos.

Major bile duct injuries were defined as the transection or significant laceration of the common hepatic duct or the common bile duct[25].

Inter-rater agreement between reviewers assessing CVS criteria in videos was quantified in terms of Cohen's *kappa*. The level of agreement was considered almost perfect ($\kappa > 0.90$), strong ($\kappa$ 0.80-0.90), moderate ($\kappa$ 0.60-0.79), weak ($\kappa$ 0.40-0.59), minimal ($\kappa$ 0.21-0.39), or no agreement ($\kappa$ 0-0.20)[26].

Categorical variables were reported with integers and frequency (%) and were compared using Fisher's exact test. Odds ratio (OR) and its 95 % confidence intervals (CI) were also reported when comparing categorical values. Normally distributed continuous variables were reported with means (±standard deviation) and compared using the two-tailed independent samples t-test whereas not normally distributed continuous variables were reported with medians (±interquartile range) and compared using the Mann-Whitney rank test. Missing data were not statistically imputed following a complete case analysis approach. Findings with *p* values <0.050 were considered statistically significant. All analyses were implemented in Python using the SciPy library[27] consisting of a package for statistical functions.

**Results**

The 5-second rule QI intervention took place on December 4, 2018. A total of 381 LC procedures were logged during the 2-year long study period. After exclusion of 6 procedures with incomplete video recordings and 32 procedures for which clinical information could not be collected, 171 and 172 LC procedures respectively performed the year before and the year after the 5-seconds rule by 17 different surgeons were included in the study. The inter-rater agreement between reviewers assessing CVS criteria in the 343 LC videos was moderate, with Cohen's *kappa* ranging from 0.72 to 0.78.

Patients' and diseases' presentations

After the implementation of the 5-second rule, significantly more patients were operated on for a previous acute cholecystitis (39.2 *vs* 51.7 % before and after the 5-second rule respectively; OR 1.66, 95 % CI 1.08 to 2.56; *P*=0.023) and had higher aspartate transaminase (22 [±10] *vs* 24 [±15] U/L; *P*=0.041) and gamma-glutamyl transferase (27 [±35] *vs* 38 [±75] U/L; *P*=0.021) levels. A comparison of patients', diseases', and operators' characteristics is shown in Table 1.

**Table 1. Comparison between the two study groups.** Categorical variables were reported with absolute numbers (frequency), normally distributed continuous variables with means (standard deviation) and not normally distributed continuous variables with medians (interquartile range). Significant differences are presented in bold.

|  | **Before (number: 171)** | **After (number: 172)** | **OR** | **95 % CI** | *p* value |
|---|---|---|---|---|---|
| **Patients' presentation** | | | | | |
| Female patients | 104 (60.8 %) | 101 (58.7 %) | 0.92 | 0.60 to 1.41 | 0.741 |
| Age (years) | 54 (±27) | 58 (±27) | NA | NA | 0.411 |
| BMI | 27 (±8) | 28 (±7) | NA | NA | 0.819 |
| ASA | 2 (±1) | 2 (±1) | NA | NA | 0.796 |
| Elective | 166 (97.1 %) | 167 (97.0 %) | 1.00 | 0.29 to 3.54 | 1.000 |
| Previous UGI surgery | 13 (7.6 %) | 14 (8.1 %) | 1.08 | 0.49 to 2.36 | 1.000 |
| Previous ERCP | 23 (13.5 %) | 27 (15.7 %) | 1.20 | 0.66 to 2.19 | 0.647 |
| Previous percutaneous drain | 8 (4.7 %) | 15 (8.7 %) | 1.95 | 0.8 to 4.72 | 0.194 |
| **Indication for LC\*** | | | | | |
| AC | 4 (2.3 %) | 4 (2.3 %) | 0.99 | 0.24 to 4.04 | 1.000 |
| **Previous AC** | **67 (39.2 %)** | **89 (51.7 %)** | **1.66** | **1.08 to 2.56** | **0.023** |
| Symptomatic cholelithiasis | 80 (46.8 %) | 68 (39.5 %) | 0.74 | 0.48 to 1.14 | 0.192 |
| Previous choledocholithiasis | 23 (13.5 %) | 25 (14.5 %) | 1.09 | 0.59 to 2.02 | 0.877 |
| Acute biliary pancreatitis | 1 (0.6 %) | 1 (0.6 %) | 0.99 | 0.06 to 16.02 | 1.000 |
| Previous pancreatitis | 27 (15.8 %) | 19 (11.1 %) | 0.66 | 0.35 to 1.24 | 0.209 |
| **Laboratory findings** | | | | | |
| Leukocytes ($10^9$/L) | 7,109 (±2,865) | 6,760 (±3,165) | NA | NA | 0.294 |
| ALT (U/L) | 23 (±17) | 27 (23) | NA | NA | 0.109 |
| **AST (U/L)** | **22 (±10)** | **24 (±15)** | **NA** | **NA** | **0.041** |
| ALP (U/L) | 76 (±31) | 77 (±34) | NA | NA | 0.180 |
| **GGT (U/L)** | **27 (±35)** | **38 (±75)** | **NA** | **NA** | **0.021** |
| CRP (mg/L) | 4 (±5.4) | 4 (±10) | NA | NA | 0.132 |
| TB (µmol/L) | 9.92 (±6.33) | 9.06 (±6.16) | NA | NA | 0.620 |
| DB (µmol/L) | 3.08 (±2.22) | 3.42 (±3.08) | NA | NA | 0.072 |
| IB (µmol/L) | 6.16 (±5.30) | 6.33 (±4.62) | NA | NA | 0.842 |
| **Surgical operators** | | | | | |
| Senior residents\*\* | 106 (62.0 %) | 108 (62.8 %) | 1.03 | 0.67 to 1.60 | 0.911 |

\*Patients may have more than one indication for surgery.
\*\*The remaining procedures were all performed by consultant surgeons.
OR: odds ratio; CI: confidence interval; BMI: body mass index; ASA: American Society of Anaesthesiologist score; UGI: upper gastrointestinal surgery; AC: acute cholecystitis; ERCP: endoscopic retrograde cholangiopancreatography; ALT: alanine aminotransferase; AST: aspartate transaminase; ALP: alkaline phosphatase; GGT: gamma-glutamyl transferase; CRP: C-reactive protein; TB: total bilirubin; DB: direct/conjugated bilirubin; ID: indirect/unconjugated bilirubin.

LC procedures

Overall, surgical operators bailed out from conventional LCs more frequently after the implementation of the 5-second rule (14 [8.2 %] *vs* 27 [15.7 %] before and after the 5-second rule respectively; OR 2.09, 95 % CI 1.05 to 4.14; $P$=0.045). Specifically, a greater number of surgeons switched to a fundus-first cholecystectomy (7 [4.1 %] *vs* 22 [12.8 %]; OR 3.44, 95 % CI 1.43 to 8.28; $P$=0.006) rather than opting for a subtotal cholecystectomy (2 [1.8 %] *vs* 3 [1.7 %]; OR 1.50, 95 % CI 0.25 to 9.09; $P$=1.000) or to abort the LC (5 [2.9 %] *vs* 2 [1.2 %]; OR 0.39, 95 % CI 0.07 to 2.04; $P$=0.283). The rate of IOC was comparable between the study groups (13 [7.6 %] *vs* 11 [6.4 %]; OR 0.83, 95 % CI 0.36 to 1.91; $P$=0.679).

Excluding bailout procedures, the CVS achievement rate increased from 25 of 157 (15.9 %) procedures to 64 of 145 procedures (44.1 %) after the implementation of the 5-second rule (OR 4.17, 95 % CI 2.43 to 7.15; $P$<0.001). The greatest improvements were noted for the achievement of the cystic plate criterion (42 [26.8 %] *vs* 93 [64.1 %] before and after the 5-second rule respectively; OR 4.90, 95 % CI 3.00 to 7.99; $P$<0.001). However, the hepatocystic triangle criterion (51 [32.5 %] *vs* 86 [59.3 %]; OR 3.03, 95 % CI 1.89 to 4.85; $P$<0.001) and the 2-structure criterion (85 [54.1 %] *vs* 104 [71.7 %]; OR 2.15, 95 % CI 1.33 to 3.47; $P$=0.002) were also significantly more achieved after the implementation of the intraoperative time out. The evolution of the CVS achievement rate over the 2-year study period can be seen in Figure 2.

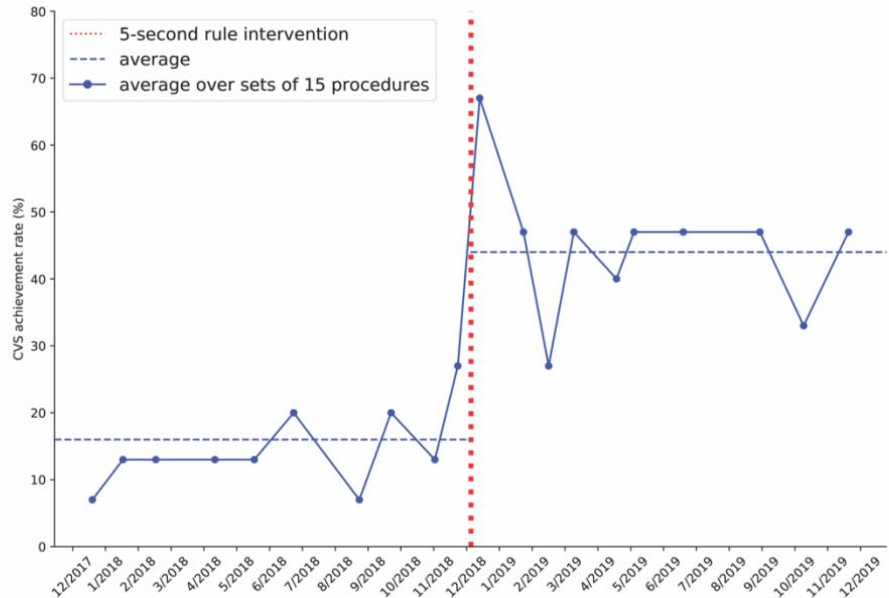

**Figure 2. Evolution of the CVS achievement rate over the 2-year study.** The vertical dashed red line marks the day of the 5-second rule QI intervention; each dot on the continuous blue line represents the CVS achievement rate averaged over sets of 15 procedures and the dashed blue line denotes the average value before and after the 5-second rule CVS achievement rate.

A sub-analysis on the CVS achievement rate of consultants and senior residents is shown Figure 3.

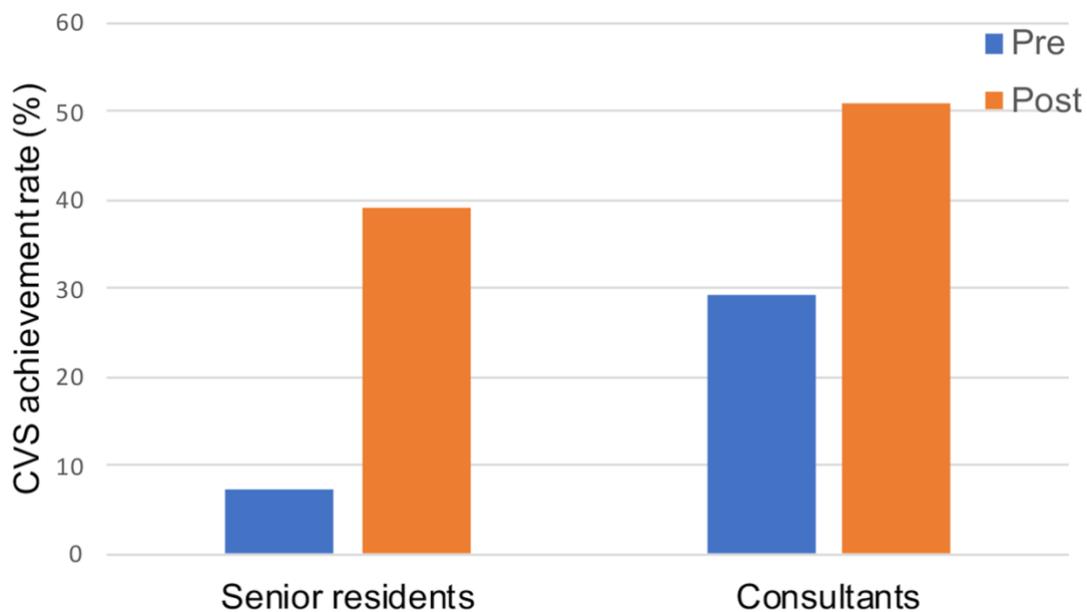

**Figure 3. CVS achievement rate of consultants and senior residents.** After the implementation of the 5-second rule, the CVS achievement rate significantly improved among consultants (17 [29.3 %] *vs* 27 [50.9 %] before and after the 5-second rule respectively; OR 2.50, 95 % CI 1.15 to 5.47; *P*=0.032) and senior residents (7 [7.2 %] *vs* 34 [39.1 %]; OR 8.25, 95 % CI 3.42 to 19.91; *P*<0.001).

Finally, operating times were comparable (46:11 [±34:36] *vs* 53:03 [±36:50] minutes before and after the 5-second rule respectively; *P*=0.088). However, surgeons spent more time before applying clips in the hepatocystic triangle after the implementation of the 5-second rule (17:26 [±16:46] *vs* 23:12 [±17:16] minutes; *P*=0.007).

CVS reporting

All but 27 missing operative reports were retrieved. After exclusion of operative reports of bailout procedures, 151 and 118 reports respectively from before and after the 5-second rule were analysed. The vast majority of the operative reports in both study groups described the dissection of Calot's triangle (148 of 151 [98.0 %] *vs* 116 of 118 reports [98.3 %] before and after the 5-second rule respectively; OR 1.18, 95 % CI 0.19 to 7.15; *P=1*.000) and the identification of the cystic artery and the cystic duct, namely the 2-structure criterion of CVS (146 [96.7 %] *vs* 115 [97.5 %]; OR 1.31, 95 % CI 0.31 to 5.61; *P=1*.000). However, explicit CVS reporting (2 [1.3 %] *vs* 34 [28.8 %]; OR 30.15, 95 % CI 7.07 to 128.68; *P*<0.001), description of the hepatocystic triangle criterion (0 *vs* 26 [22.0 %]; *P*<0.001), and the cystic plate criterion (0 *vs* 26 [22.0 %]; *P*<0.001) increased significantly after the 5-second rule.

LC clinical outcomes

No major BDIs nor fatalities occurred during the study period. Length of hospital stay (0 [±2] *vs* 0 [±2] days before and after the 5-second rule respectively; *P*=0.970), readmission (1 [0.6 %] *vs* 3 [1.7 %]; OR 3.02, 95 % CI 0.31 to 29.30; *P*=0.623), and reintervention (1 [0.58 %] *vs* 1 [0.58 %]; OR 0.99, 95 % CI 0.06 to 16.02; *P*=1.000) rates were comparable among the study groups. Morbidity was also comparable (2 [1.2 %] *vs* 4 [2.3 %]; OR 2.01, 95 % CI 0.36 to 11.13; *P*=0.685), with an overall bile leak rate of 5 in 343 LCs [1.5 %]. Details on patients who experienced morbidity in the study period can be found in Table 2.

**Table 2. Patients who experienced adverse events during the study.**

| Patient | Study group | Indication for LC | CVS or bailout | Operating time (hh:mm:ss) | Adverse event (if BDI, Strasberg type[7]) | Dindo-Clavien grade[28] | Treatment |
|---|---|---|---|---|---|---|---|
| 1 | Before | Previous AC, previous pancreatitis | Not achieved | 01:30:00 | BDI (A) | IIIb | Laparoscopic suture of Luschka duct |
| 2 | Before | Previous AC, previous pancreatitis | Converted | 02:52:00 | BDI (A) | IIIb | Percutaneous drain, ERCP stenting |
| 3 | After | Previous AC, previous pancreatitis | Not achieved | 01:00:00 | BDI (A) | IIIb | Percutaneous drain |
| 4 | After | Symptomatic cholelithiasis | Not achieved | 00:32:00 | BDI (C) | IIIb | Percutaneous drain, laparoscopic lavage |
| 5 | After | Previous AC, previous pancreatitis, previous cholelithiasis | Fundus-first LC | 02:00:00 | Intra-abdominal hematoma | IIIb | Percutaneous drain |
| 6 | After | Previous AC, previous cholelithiasis | Subtotal LC | 03:32:00 | BDI (A) | IIIb | Percutaneous drain, ERCP stenting |

LC: laparoscopic cholecystectomy; CVS: critical view of safety; BDI: bile duct injury; AC: acute cholecystitis; ERCP: endoscopic retrograde cholangiopancreatography.

**Discussion**

The present quality improvement (QI) study evaluated the impact of the 5-second rule intraoperative cognitive aid in a series of 343 LCs using a validated protocol for the video assessment of the CVS[21]. In this study, implementing a 5-second long intraoperative time out to verify the CVS before dividing the cystic duct led to an approximately threefold increase in the average CVS achievement rate (Figure 2). Improvements were consistent across the 3 CVS criteria and across surgical operators with different levels of experience (Figure 3). In addition, the more frequent decision to bail out, the longer time taken to carefully dissect the hepatocystic triangle, and the increased rate of postoperative CVS reporting after implementation of the 5-second rule suggest an increased awareness towards CVS principles and the so-called "Culture of Safety in Cholecystectomy"[29].

These findings are in line with the results of a previous pilot study evaluating the effect of comprehensive education and a CVS time out in a series of 101 LC cases performed over a 5-month study period[30] and confirm the recommendations[1, 30] on the importance of performing an intraoperative pause to implement best practices in LC.

The present findings hint at a series of observations which deserve further research. With respect to overall LC safety, the fact that CVS-aware surgeons seem to have a lower threshold to bail out from difficult procedures may be positive[20]. However, in the present series surgeons have more often opted for a fundus-first cholecystectomy rather than a subtotal cholecystectomy. The right threshold for bailing out and the best alternative to a standard LC are a matter of fervid scientific debate[1, 31-33] and future studies should be designed to address these points.

With respect to the CVS, the achievement rate before the 5-second rule intervention was impressively low, an observation also reported by a recent work[14]. Soon after the QI intervention, the rate of CVS achievement peaked to almost 70 % but then stabilised at approximately 45 % at 1 year (Figure 2). The peak suggests the operating know-how to correctly implement CVS. On the other hand, the reduction in CVS achievement over time may indicate a decline in the application of the 5-second rule. Overall, the inconsistent implementation of CVS together with its subjective assessment and reporting[12-14] might hamper efforts to prevent BDIs in LC. These limitations have traditionally been addressed mostly through education. To better penetrate surgical practices, cholecystectomy-specific intraoperative checklists[35] and stepwise guidelines for difficult cases[36] have also been developed. Today, surgeons and computer scientists from various institutions including ours have teamed up and embarked on a series of surgical data science studies[37–39] which will hopefully lead to the development of context-aware[40] computer vision solutions capable of automatically reminding, univocally assessing[23] and objectively reporting safety measures such as the CVS[41].

Finally, this study has some limitations. From a methodological standpoint, the use of VBA is both a merit and a drawback. VBA is increasingly recognised as a valuable approach to objectively study intraoperative surgical performance[17, 40], especially when operative reports are known to be unreliable as in the case of the CVS[12, 41]. However, the potential Hawthorne effect (i.e. behaviour changes due to the subjects' awareness of being observed, also known as the clinical trial effect) impedes to conclude to what extent the improvements in CVS achievement rates were either due to the operators' awareness of being recorded or to the 5-seconds rule. This is especially true since the application of the 5-second rule by surgeons was not appropriately studied through in-person observations in the operating room. The absence of an observer in the operating room and the fact that our institution systematically records surgical procedures for QI and education certainly decreases the Hawthorne effect. However, the extent to which this is the case is unknown[44] so far. From a clinical perspective, the study was not adequately powered to spot a difference in clinical outcomes. This is a limitation common to most studies on BDIs in LC, as an impractical large number of patients would be necessary to spot a statistically significant difference in the incidence of such a rare adverse event[19]. However, the CVS achievement rate, which is the primary endpoint of this study, is considered to be an acceptable process measure of LC safety given the widely accepted correlation between this critical view and the prevention of major BDIs. Finally, as the vast majority of LC cases included were performed in the elective setting due to video recording constraints, further research is required to confirm the impact of the 5-second rule time out in the acute setting.

In conclusion, the 5-second rule time out significantly increased the CVS achievement rate. Overall, the findings of the present article offer a practical strategy to improve the uptake of safety principles in LC. Future studies should explore the use of intraoperative cognitive aids and develop solutions to sustain their positive effect over time.

**Acknowledgments**

The authors would like to acknowledge Guy Temporal for his assistance with English proofreading, Deepak Alapatt for his recommendations on statistical analyses, and Armine Vardazaryan for her help in the organisation of the video analyses.

The present study was not pre-registered in an independent institutional registry.